\documentclass[twocolumn, aps, pra]{revtex4-1}

\usepackage{graphicx}
\usepackage{dsfont}
\usepackage{amsfonts}
\usepackage{xcolor}
\usepackage{physics}
\usepackage{ulem}
\definecolor{amber}{rgb}{1.0, 0.75, 0.0}
\definecolor{babyblue}{rgb}{0.54, 0.81, 0.94}
\definecolor{ao(english)}{rgb}{0.0, 0.5, 0.0}
\RequirePackage{hyperref}
\hypersetup{
    colorlinks=true,
    linkcolor=blue,
    citecolor=blue,
    urlcolor=blue,
}
\date{\today}

\begin{document}

\title{Finite-time collapse process and non-local correlations are incompatible with non-signaling theories}

\author{M. G. M. Moreno}
\email[]{marcosgeorge.mmf@gmail.com}
\author{Alejandro Fonseca}
\affiliation{Departamento de F\'isica, Universidade Federal de Pernambuco. Recife 50670-901, PE, Brazil.}
\author{M\'arcio M. Cunha}
\affiliation{Departamento de Matem\'atica, Universidade Federal de Pernambuco. Recife 50670-901, PE, Brazil.\\
Departamento de F\'isica, Universidade Federal Rural de Pernambuco. Recife 52171-900, PE, Brazil.}
%
%
\begin{abstract}
We propose a hidden variable analysis of collapse dynamics
in which the state's reduction process may take a finite time $\delta t$. A full characterization of the model is given for the 
case of black boxes. By introducing nonlocal perfect correlations to a 
two black-boxes scenario, it is shown that in order to avoid faster than light communication, the reduction 
time associated to the system must be strictly null. Furthermore we prove that the result above holds even when there is a time window between the choice of both part's inputs. Our results represent a new evidence of the instantaneous nature of the wave function collapse process which could have implications in foundations of quantum mechanics and information science.
\end{abstract}

\maketitle

\section{Introduction}

After almost a century of Quantum Theory, one of its fundamental postulates represents yet an issue: the measurement problem. This widely discussed phenomenon, in principle forbidden by the linearity of Schr\"odinger's equation, motivated many works aiming to present an explanation on the mechanism leading the system, in a probabilistic way to the collapsed state, in perfect harmony with Born's rule and L\"{u}ders postulate. From the perspective of the Copenhagen interpretation those problems are solved by taking them as fundamental postulates \cite{bohr1928quantum}. Following the \textit{Many World} interpretation, all possible outcomes from a measurement coexist in different universes \cite{everett1957relative}, avoiding the necessity of L\"{u}ders postulate nevertheless it doesn't explain the mechanism behind Born's rule. Decoherence \cite{zurek2003decoherence} approaches the problem by showing that a quantum system in contact with a environment should swiftly reach a classic statistical distribution, nonetheless closed systems remain a problem. Among the attempts to solve the measurement problem, the most plausible explanation is that Quantum Mechanics is an approximated theory, in the sense that there must a nonlinear equation that rules the dynamics of all systems which in the microscopic limit becomes approximately linear in accordance with Schr\"odinger equation, after all no microscopic experiment ever indicated Quantum Theory to be wrong. Such a theory should also be stochastic, since any nonlinear deterministic extra term in Schr\"odinger's equation leads to signaling \cite{gisin1990weinberg}.
That approach motivated many collapse models introduced in the last few decades, as well as several experiments were proposed to investigate the features of the wave function collapse \cite{bahrami2014proposal,diosi2015testing, goldwater2016testing, nimmrichter2014optomechanical, genoni2016unravelling, bilardello2017collapse, vinante2016upper,vinante2017improved}. For a complete review on the current state of the area see \cite{bassi2013models}.

If collapse comes from a dynamical process, a fundamental question should be addressed: \textit{once initiated, how long, in average, does the process of collapse take to be accomplished?} The mean time of the collapse may bring a deep insight on the dynamics that rules it. This problem, which has already been considered and experimental tests to verify the duration of the collapse have been proposed \cite{parisio2011estimating,moreno2013investigation,moreira2018toward}, will be the central issue of this work.

In order to investigate the mean time associated with the collapse process, the present work brings into play a quite general approach: a hidden variables model, in a Device Independent (DI) scenario. A hidden variables model allows for the contemplation of possible consequences and effects of unknown parameters, and its dynamics, that may be playing some rule on a given problem. Under this approach, we are particularly interested into hypotheses on the set of hidden variable which can generate appreciable differences in the outputs of some experiment. This method is well know by the seminal work of J. Bell \cite{Bell1964}. Recently Bedingham employed this concept to perform a link between the collapse dynamics and the Bohmian mechanics \cite{1751-8121-44-27-275303} (see also \cite{1751-8121-44-47-478001}). On the other hand, the DI certification program provides robust results, for it only relies on the statistics of a given experiment and it is not necessary to make any extra assumption on the system to be tested. A system under the DI procedure is treated as an assortment of boxes equipped with buttons which after being pushed produce one out of an array of outcomes, in general different in each run of the experiment. Associated with hidden variable models, the DI idea has been of remarkable importance on protocols of certification. For an introduction to the subject, we refer the reader to \cite{scarani2012device}. At the end of the day, from the set of frequencies it is possible to infer underlying properties of the whole system. In this paper, by using nonlocally correlated systems and applying the ideas above we obtain general results that are independent of Quantum Mechanics, however, they lead to very remarkable consequences in its context.

The paper is organized as follows: In section II we show the usual hidden variable (HV) model in the context of non-local correlations. After in section III we introduce a HV model to treat the problem of a single system subject to an arbitrary collapse dynamics. Section IV is devoted to make a connection between nonlocal correlations and collapse dynamics under the HV models overlook, and we present some implications of the collapse dynamics. In section V we expose our main conclusions.

\section{Hidden variables and non-local correlations}

The most known application of hidden variables models is perhaps the definition of local correlations, associated with the derivation of Bell's inequalities. In this problem it is investigated the statistical behavior $P(a,b|x,y)$ of two separated parts in which inputs are performed, $x$ in one and $y$ in the other side, generating outputs $a$ and $b$ respectively. We can always write:
\begin{equation}
\label{HV1}
P(a,b|x,y)=\int_{\Lambda}d\lambda\zeta(\lambda|x,y)P(a,b|x,y,\lambda).
\end{equation}
In the above expression, $\lambda$ represents variable(s) which are sampled from a set $\Lambda$, following a distribution $\zeta (\lambda|x,y)$, responsible for the probability of the system, and yet out of the reach for the experimenter.

Equation (\ref{HV1}) represents the basic assumption behind hidden variable models, however one can always add extra hypothesis on them. For instance, one can assume that superluminal communication between parts is forbidden, i. e. the non-signaling assumption, in this case the marginal probabilities of each part should be independent of what happens in the other:
\begin{eqnarray}
\nonumber
P(a|x,y,\lambda)=P(a|x,\lambda)\\
\nonumber
P(b|x,y,\lambda)=P(b|y,\lambda),
\end{eqnarray}
then:
\begin{equation}
P(a,b|x,y,\lambda)=P(a|x,\lambda)\cdot P(b|y,\lambda).
\end{equation}
We can go further and assume that all correlations come from the $\lambda$ variables and that their distribution is well defined despite the inputs $x$ and $y$, i. e., $\zeta(\lambda|x,y)=\zeta(\lambda)$, which is known as the measurement independence assumption. Moreover note that the inputs $x$ and $y$ do not depend on the set of hidden variables (free will assumption). Considering this, correlations described by a local hidden variables model may be written as:
\begin{equation}
\label{LV}
P(a,b|x,y)=\int_{\Lambda}d\lambda~\zeta(\lambda) P(a|x,\lambda)P(b|y,\lambda),
\end{equation}
attainable by any classically correlated composed system.

This example illustrates the power of hidden variables assumptions and how to handle them in order to get valuable information on the system under consideration.

\section{Hidden Variables and Collapse}

Imagine Alice receives a closed box containing a flipped coin. Right before Alice looks inside the box, she would say that the probability of getting either heads or tails is one half (assuming a faithful coin). If Alice finds that the output was tails (heads) any further observation of the same coin will yield the output tails (heads) with probability one. We may say that Alice's system collapses after it is measured. So far the problem seems very trivial: the position of the coin is well defined from the moment the system was created, and the act of looking to the coin just means to learn the value of some unknown well defined variable. This is not always the case, for instance one may consider that instead of a coin, there is an electron inside the box prepared in a spin state $(\ket{0}+\ket{1})/\sqrt{2}$, for $\ket{0}$ and $\ket{1}$ representing the eigenstates of $\hat{\sigma}_z$. When Alice carries out a spin measurement in $z$ direction, quantum theory states that there are not \textit{a priori} established variables hidden from Alice defining the corresponding outcomes. In contrast with the former case, now a physical process is expected to take place, leading to the final outputs. This is what we mean by \textit{collapse}. 

Following the above scenario, now we consider Alice receiving a box, on which she can provide an input $x$ from a set of inputs $\mathcal{X}$ that generates some output $a\in\mathcal{A}$, with a well defined probability. Before making any further assumption, it is useful to introduce a formal distinction between the two classes of inputs that Alice may provide to her box. On one hand, we have inputs leading to  outcomes in a non-deterministic way -hereafter \textit{collapse triggering} operations (CT). On the other hand, operations conducting to deterministic outputs, defined here as \textit{non collapse triggering} operations (NCT). Hence it is possible to divide the set of inputs $\mathcal{X}$ in two parts: $\mathcal{X}=\mathcal{X}_{CT}\cup\mathcal{X}_{NCT}$.

In order to contemplate any possible collapse dynamics (consequence of a CT input) and its effect on the description of an arbitrary system, we propose a hidden variable approach similar to that introduced in previous section. Let state some assumptions: (i) the collapse is triggered by an input $x\in\mathcal{X}_{CT}$ on the box at an instant $\tau$ which returns some output $a\in \mathcal{A}$, (ii) the system takes a time $\delta t_a$ to collapse (i.e. to generate an output), (iii) the probability of obtaining the output ``$a$" as a consequence of the first input $x$ is known to be $P_0(a|x)$, and (iv) the collapse time $\delta t_a$ depends in a non trivial way on the probability of its output $P_0(a|x)$. Considering that, the most general expression describing the probability of an output $a'$ given a second input $x$ in a posterior time $\tau\leq t\leq \tau+\delta t$, is:
\begin{equation}
\label{HV2}
P(a'|x;t>\tau)=\int_{\Gamma }d\gamma(t)\chi(\gamma(t)|x)P(a'|x;\gamma(t)),
\end{equation}
here $\Gamma$, $\chi$ and $\gamma$ play the same role as $\Lambda$, $\zeta$ and $\lambda$ in equation \ref{HV1}, respectively. This model encompasses any possible description behind the phenomenon of collapse. In fact we could make $\chi(\gamma(t)|x)=\delta[\gamma(t)-\hat{\rho}(t)]$, where $\delta[.]$ is the Dirac's delta and $\hat{\rho}(t)$ is a density operator, and considering the POVM $\hat{M}_x=\lbrace \hat{E}_a^x|a\in \mathcal{A}\rbrace$, and $P(a'|x;\gamma(t))=\tr(\gamma(t)\cdot\hat{E}_{a'}^x)$, then we have:
\begin{equation}
P(a'|x;t>\tau)=\tr(\hat{\rho}(t)\cdot\hat{E}_{a'}^x),
\end{equation}
which corresponds to the standard formulation in quantum mechanics \cite{scarani2012device}.

After the largest among the collapse times ($t\geq \tau+\delta t^*$), where $\delta t^*=\max\lbrace\delta t_a\rbrace_{a\in\mathcal{A}}$, we expect the box to evolve in such a way that $P(a'|x;t\geq\tau+\delta t)=\delta_{a,a'}$, where $\delta_{a,a'}$ is the Kronecker's delta, and $a$ represents the first output. Without loss of generality we can divide the set of variables $\Gamma$ into subsets $\Gamma_a$, each containing all possible $\gamma(t)$ leading to every output ``$a$". As we know \textit{a priori} that the result ``$a$" should be obtained with probability $P_0(a|x)$, then we can write:
\begin{equation*}
P(a'|x;t)=\sum_{a\in\mathcal{A}}P_0(a|x)\int_{\Gamma_a }d\gamma(t)\chi_a(\gamma(t)|x)P(a'|x;\gamma(t)),
\end{equation*}
for $\tau\leq t\leq \tau+\delta t^*$. To gain some insight on this particular step, it is possible to consider the one-dimensional random walker, which after $n$ steps has a probability $P(j|n)$ of being found in the position $j$. There may exist several possible paths leading to this configuration,
thus one can assemble all these paths together in the set $\Gamma_j$ and argue that with probability $P(j|n)$ a path from this set is sorted out. Also notice that no knowledge from the outputs to be obtained is required to conceive the existence of this partition, only the assumption that one of the possible results will happen.

We can simplify the above equation, by defining the functions $f_{aa'}(t)$:
\begin{equation}
f_{aa'}(t)=\int_{\Gamma_a }d\gamma\chi(\gamma(t)|x)P(a'|x;\gamma(t)),
\end{equation}
which should respect the following bounds:
\begin{eqnarray}
\label{bc1}
\nonumber
&f_{aa'}(\tau) =P_0(a'|x),&\\
&f_{aa'}(t'\geq\tau +\delta t_a) = \delta_{a,a'},&\\
\nonumber
&\sum_{a'} f_{aa'}(t'\geq\tau) = 1.&
\end{eqnarray}
The first condition is related to the initial probability distribution of the box, the second sets the final configuration after the collapse and the last one ensures that normalization is satisfied.

Thus using these new functions, we have that:
\begin{equation}
\label{col0}
P(a'|x;t\geq \tau)=\sum_{a} f_{aa'}(t)P_0(a|x).
\end{equation}
Note that we have not considered any specific dynamics so that this result remains as general as possible. Furthermore, the collapse time intervals $\delta t_a$ can be taken as zero or finite without loss of generality.

Equation (\ref{col0}) and relations (\ref{bc1}) lead to the following result: if the second input $x$ is performed at an instant $t\geq\delta t^*$, the output $a$ will occur with probability $P_0(a|x)$, however if $t\leq\delta \tilde{t}$, for $\delta\tilde{t}=\min\lbrace\delta t_a \rbrace_{a\in\mathcal A}$, then the probability distributions will depend on the functions $f_{aa'}(t)$. In particular, we can consider the quantity:
\begin{equation}
P(a|x;\tau\leq t\leq \delta\tilde{t})-P_0(a|x),
\end{equation}
which can be experimentally assessed. For a dichotomic system, the only condition that allows it to be zero would be $P(a|x)=\frac{1}{2}$, otherwise this quantity is non-vanishing.

\section{Hidden Variables, Non-locality and collapse}
\label{3}
Following the previous reasoning, an extension to the case of two separated boxes sharing some correlation is presented. Two balls, one black and the other white, are randomly placed into Alice and Bob's boxes respectively. If Alice looks inside her box and learns the colors of her ball, then due to the correlation they shared, she also learns that of Bob. Setting $a=\{0,1\}$ as the color of the ball and $x=1$ to the act of measuring it, the first time the input $x=1$ is provided, the output $a$ is obtained with probability $P_0(a|x=1)$, however any further ``color measurement" will yield the output $a'$ with probability $P(a'|x=1)=\delta_{a,a'}$, given that we are leading with a locally-correlated system. The same behavior is observed in Bob's box.
Like the first example, in this case there are variables that could give a complete description of the box at first, but are not revealed to the parts. Thus the collapse represents only the knowledge of some hidden variable. No dynamical process is expected here, for there is no evidence of physical changes. To observe some collapse dynamics as discussed above, one must look for correlations that cannot be represented by a local hidden variables model (eq. \ref{LV}), where the collapse represents a physical transformation in both parts.





With this in mind, assume we have two arbitrarily separated parts, Alice and Bob as usual, both in inertial reference frames, possessing nonlocally correlated boxes. 
Alice can provide either an input or $x=0\in\mathcal{X}_{NCT}$ or $x=1\in\mathcal{X}_{CT}$, such that $P(a|0)=\delta_{a,0}$, obtaining some output $a\in\mathcal{A}$, and Bob also gives either an input $y=0\in\mathcal{Y}_{NCT}$ or $y=1\in\mathcal{Y}_{CT}$ where $P(b|0)=\delta_{b,0}$ returning some output $b\in\mathcal{B}$. Furthermore, assume that the correlation is such that given $x=1$ and $y=1$, then $a=b$, and that the probability of the first measurement in the system $P_0(a,b|x=1,y=1)=P_0(a|x=1)=P_0(b|y=1)$ is known, where the equalities hold due to the correlation which forbids results where $a\neq b$. Notice that the input pairs $(x=0,y=1)$, $(x=1,y=0)$, $(x=1,y=1)$ are collapse triggering.

Suppose Alice and Bob agree that at an instant $\tau$ in Bob's watch, she decides the value for $x$, while Bob sets $y=0$. And at an instant $t'\geq\tau$, Alice provides the input $x=0$, and Bob $y=1$. In this point the natural step is to compare both scenarios $x=0$ and $x=1$, given by Alice's initial choice. 
This is a crucial aspect in our approach, for usually when the subject of collapse is tested with respect to whether it is signaling or not, only collapse triggering inputs are considered \cite{bedingham2009dynamical}. 

Whenever Alice chooses $x=0$, Bob providing $y=1$ in $t'$ will be able to observe that the outputs $b$ follow the known distribution:
\begin{eqnarray}
\label{X1}
P(b|0,1;t')=P_0(b|1).
\end{eqnarray}
Nevertheless, when she supplies $x=1$, the system as a whole starts to collapse, and according to equation (\ref{HV2}) the probability in $t'$ can be described by:
\begin{equation}
\nonumber
P(a,b|0,1;t)=\int_{\Gamma}d\gamma(t)\chi(\gamma(t)|0,1)P(a,b|0,1;\gamma(t)).
\end{equation}
Following the treatment for the single box, it is a fact that the system collapses to some output $(a,b)$ with probability $P_0(a,b|1,1)=\delta_{a,b}P_0(b|1,1)$, assuming perfect correlation:
\begin{equation}
\nonumber
P(b'|1,1;t)=\sum_{b\in\mathcal B}P_0(b|1)\int_{\Gamma_b}d\gamma(t)\chi_b(\gamma(t)|1)P(b|1,1;\gamma(t)).
\end{equation}
Here we define:
\begin{equation}
f_{bb'}(t)=\int_{\Gamma_b}d\gamma(t)\chi(\gamma(t)|1)P(b|1,1;\gamma(t)),
\end{equation}
such that:
\begin{eqnarray}
\nonumber
&f_{bb'}(\tau)=P_0(b'|y),&\\
\nonumber
&f_{bb'}(t'\geq\tau +\delta t) = \delta_{b,b'},&\\
\nonumber
&\sum_{b'} f_{bb'}(t'\geq\tau) = 1,&
\end{eqnarray}
so we can write:
\begin{equation}
\label{X2}
P(b|1,1;t') = \sum_a f_{bb'}(t')P_0(b|1).
\end{equation}
It is always possible to find distributions for which equations (\ref{X1}) and (\ref{X2}) are in agreement with non-signaling conditions if and only if $f_{00}(t')=f_{11}(t')=1$ and $f_{01}(t')=f_{10}(t')=0$ for $t'>\tau$, representing an instantaneous collapse. Otherwise Alice's choice affects Bob's statistics. This analysis suggests that only instantaneous collapses are compatible with non-signaling theories.

The results shown above are quite general in the sense that no assumptions are made on the dynamics behind the process. However this scheme does not contemplate all possible scenarios yet. Particularly, we have assumed that Alice and Bob can choose the specific time in which the inputs are delivered, which is not necessarily feasible. For instance, following Quantum Theory one cannot choose exactly the instant in which a photon is emitted nor when it will hit the detector. Now, we repeat our analysis by taking into account that Alice and Bob can only decide the time window  $\Delta t>\delta \tilde{t}$ in which the input acts happen. In addition, we consider that each of them can only perform one input: Alice deciding $x$ and Bob applying $y=1$.

When the time window starts we assume that an input takes place according to some probability distribution $g(t)$. Thus, the probability of it to happen at an instant $t'$, $\tau\leq t'\leq \tau + \Delta t$, is:
\begin{eqnarray}
\mathcal{P}(t')=\int_{\tau}^{t'}g(t)dt,
\end{eqnarray}
where $\int_{\tau}^{\tau+ \Delta t}g(t)dt=1$.

Now we are interested in what happens if the time interval between the instant in which Alice's input happens, $t_A$, and the moment Bob's input occur, (hereafter $t_B$), are smaller than $\delta \tilde{t}$. This is because if the time interval is larger, then the collapse process will be completed and the observed effect in the previous case will not play any role here. In general, we have:

\begin{eqnarray}
\nonumber
\mathcal{P}(|t_B-t_A|<\delta \tilde{t})=\int_{\tau}^{\tau+\Delta t}g(t_A)\int_{t_A-\delta \tilde{t}}^{t_A+\delta \tilde{t}}g(t_B)dt_B dt_A,
\end{eqnarray}
hereafter $\Theta=\mathcal{P}(t_B-t_A\leq\delta \tilde{t})$.

Let us investigate the partial probability distribution for Bob. Once again, when Alice decides $x=0$, Bob observes that the statistics associated to his outputs is equal to the already know distribution $P_0(b|y)$. Alternatively, when she chooses $x=1$, triggering the collapse, then Bob's probability can be described in the following way:
\textcolor{black}{\begin{multline}
\label{py1}
P(b'|y)=\left(1-\Theta\right)P_0(b'|y)+\\
 +\frac{\Theta}{\Omega}\Bigg\{\sum_{b\in\mathcal{B}} P_0(b|y)\int_{0}^{\delta \tilde{t}}f_{bb'}(t')g(t')dt'\Bigg\},
\end{multline}}
where $\Omega=\int_{0}^{\delta \tilde{t}}p(t)dt$.

The positivity of the probability distributions guarantees that equation (\ref{py1}) may be different from the a priori known distribution $P_0(b|y)$, unless $f_{ab}(t')=\delta_{a,b}$ for $t'\geq\tau$. Hence the only way to avoid signaling for any distribution is an instantaneous collapse dynamics.
\section{Conclusion}
We have addressed the question of the finiteness of collapse time for two different scenarios, on one hand a single system and on the other, a bipartite correlated one, treated as boxes in analogy to Bell scenarios. Based on our idea of collapse, the inputs are divided in two sets: collapse and non-collapse triggering. 
A hidden variables approach is employed to model an arbitrary collapse dynamics. 
For the first case we demonstrate that in principle it is possible to distinguish finite time from instantaneous collapse dynamics.

For the case of two nonlocally correlated parts, we derive some conditions necessary to such correlations do not violate non-signaling constraints during the collapse process.
The obtained result is quite general, and suggests that any collapse dynamics with finite time is incompatible with non-signaling constraints.
Our results are particularly relevant in the context of Quantum Foundations, because it can can bring some insights on the measurement problem, still an open question nowadays.

\section{Acknowledgements}
We thank Rafael Chaves and Barbara Amaral for the comments and suggestions. M\'arcio M. Cunha is supported  by FACEPE-FULBRIGHT BCT 0060-1.05/18 grant. Financial support from Conselho Nacional de Desenvolvimento Cient\'{\i}fico e Tecnol\'ogico (CNPq) through its program INCT-IQ, Coordena\c{c}\~ao de Aperfei\c{c}oamento de Pessoal de N\'{\i}vel Superior (CAPES), and Funda\c{c}\~ao de Amparo \`a Ci\^encia e Tecnologia do Estado de Pernambuco (FACEPE) is acknowledged.
\bibliography{Bib}
\end{document}